\documentclass[nohyper,12pt,letterpaper]{JHEP3}
\usepackage[dvips]{epsfig}
\usepackage{amsfonts,amssymb}

\newcommand{\be}{\begin{equation}}
\newcommand{\ee}{\end{equation}}
\newcommand{\ben}{\begin{displaymath}}
\newcommand{\een}{\end{displaymath}}
\newcommand{\bea}{\begin{eqnarray}}
\newcommand{\eea}{\end{eqnarray}}
\newcommand{\bean}{\begin{eqnarray*}}
\newcommand{\eean}{\end{eqnarray*}}

\newcommand{\ads}[1]{\mbox{${AdS}_{#1}$}}

\newcommand{\eg}{{\it e.g.}}
\newcommand{\ie}{{\it i.e.}}

\newcommand{\beq}{\begin{equation}}
\newcommand{\eeq}{\end{equation}}
\newcommand{\beqr}{\begin{displaymath}}
\newcommand{\eeqr}{\end{displaymath}}
\newcommand{\beqa}{\begin{eqnarray}}
\newcommand{\eeqa}{\end{eqnarray}}
\newcommand{\beqar}{\begin{eqnarray*}}
\newcommand{\eeqar}{\end{eqnarray*}}
\newcommand{\cN}{{\cal N}}
\newcommand{\cD}{{\cal D}}
\newcommand{\cO}{{\cal O}}
\newcommand{\cF}{{\cal F}}

\newcommand{\half}{\ensuremath{\frac{1}{2}}}

\newcommand{\N}[1]{\ensuremath{\cN=#1}}

\newcommand{\ints}{\ensuremath{\int\! d^d\!x\, dt\,}}
\newcommand{\bmu}{\ensuremath{\bar{\mu}}}
\newcommand{\bnu}{\ensuremath{\bar{\nu}}}

\newcommand{\cC}{\mathcal{C}}

\title{\LARGE Renormalization group flow of reduced string actions}

\author{Martin Kruczenski \\
        Department of Physics, Princeton University \\
        Princeton, NJ 08544.

E-mail: \email{martink@princeton.edu}}

\abstract{It has been argued that certain reduced actions play a role in AdS/CFT when
comparing fast moving strings to long single trace operators in gauge theories.

 Such actions arise in two ways: as a limit of the string action and as a
description of long single trace field theory operators.
They are non-relativistic sigma models with the target space usually 
being a K\"ahler manifold.
 They are non-renormalizable and need a cut-off in the wave-length. 
If the total spin (or charge) contained in a minimal wavelength is large compared to one, 
the system behaves semiclassically and an expansion in loops is meaningful.

 In this paper we apply the renormalization group procedure to such actions and find, at one-loop, 
that the K\"ahler potential flows in the infrared to a K\"ahler-Einstein one. 

Therefore, in this context, the anomalous dimensions of long operators 
are determined by a fixed point. This suggests that certain features of 
the large N-limit might be independent of the detailed properties of a gauge theory.
}


\begin{document}

\section{Introduction}

 It has since long been suspected that four dimensional confining theories have a dual string description
in the large-N limit~\cite{largeN}. Relatively recently a precise example of a duality between a gauge 
theory and a string theory was established through the AdS/CFT correspondence~\cite{malda}. 
 Following that, Berenstein, Maldacena and Nastase~\cite{Berenstein:2002jq} show a matching
between certain operators in the boundary theory and excited strings in the bulk. Such relation 
turned out to be part of a more general relation between semi-classical  strings in the bulk~\cite{GKP} 
and certain operators in the boundary \footnote{See 
the recent reviews~\cite{Tseytlin:2003ii,Tseytlin:2004xa,Tseytlin:2004cj} for a summary  
with a complete set of references.}. In another development, Minahan and Zarembo~\cite{Minahan:2002ve} observed 
that the one-loop anomalous dimension of operators composed of scalars in \N{4} SYM theory follows from solving and 
integrable spin chain\footnote{In QCD the relation between spin chains and anomalous dimensions had already been noted 
in~\cite{Braun:1998id}.}. 
This allowed the authors of~\cite{Beisert:2003ea,Beisert:2003xu} to make a much more
detailed comparison between particular string solutions and operators in the gauge theory. 

 It was later suggested \cite{Kruczenski:2003gt} that a classical sigma model action follows from 
considering the low energy excitations of the spin chain. It turned out that such action agrees with a 
particular limit of the sigma model action that describes the propagation of the strings in the bulk. This 
idea was extended to other sectors including fermions and open strings in 
\cite{Dimov:2004qv,Hernandez:2004uw,Ryang:2004tq,Dimov:2004xi,Stefanski:2004cw,Kruczenski:2004cn,
Ideguchi:2004wm,Ryang:2004pu,Hernandez:2004kr,Bellucci:2004qr,Susaki:2004tg,open}
and to two loops in \cite{Kruczenski:2004kw}. In \cite{Kazakov:2004qf,Kazakov:2004nh} the relation to 
the Bethe ansatz approach was clarified. Further, similar possibilities were found in certain subsectors
of QCD \cite{Ferretti:2004ba}. A different but related approach to 
the relation between the string and Yang-Mills operators has been pursued 
in \cite{Mikhailov:2004qf,Mikhailov:2004xw,Mikhailov:2004au} where a geometric description in 
terms of light like world-sheets was found. In that respect see also \cite{Gorsky:2003nq}.
 A similar picture seems to appear also in other sector \cite{Kruczenski:2004wg}. 

 Recently a large class of new examples of the AdS/CFT correspondence was found.
They are of the form $\ads{5}\times X_5$ where $X_5$ is a Sasaki-Einstein manifold. 
In the new examples, $X_5$ is one of the so called $Y^{p,q}$ and $L^{p,q|r}$ manifolds whose metrics where
found in \cite{Gauntlett:2004yd,Gauntlett:2004hh} and \cite{Cvetic:2005ft}, 
and the dual gauge theories in \cite{Benvenuti:2004dy} and \cite{Benvenuti:2005ja}.

 It was natural to see if the methods of the reduced action could be applied to such case. This was
considered in \cite{BK} for the $Y^{p,q}$ case. The field theories are strongly coupled and therefore
no field theory calculations were possible. However it was suspected that the string action only 
captures generic properties of the operators. That was partially confirmed by using a simplified model
for the operators which, in the classical limit, was described by an action similar (but not equal) to the
action derived from the string side. It was further argued that in the infrared, the action of the 
simplified model should flow to the one obtained from the string side. This was based on the fact that
the metric in the string side satisfied the Einstein equations (with cosmological constant) and the idea 
that the Einstein equation was a natural candidate for an equation determining an infrared fixed point of 
the action.
  
 In this paper we analyze this problem for a class of actions that depend on a K\"ahler potential in an
$n$ complex dimensional manifold. They include most of the known examples of reduced actions.

 By doing a one-loop computation we confirm that the reduced action obtained from the string side is at
a fixed point. However for an action to flow towards that fixed point, some conditions have to be met
which do not seem to be satisfied by the simplified model of \cite{BK}. This means that, although 
the basic idea is correct, in the sense that a large class of models flow to the one of interest, 
the particular one considered in \cite{BK} does not seem to be in that class. It would be interesting
to see if there is a simple model that actually have the correct properties. 

 Before proceeding we should note that, in the $SU(2)$ case, loop corrections were studied in several 
papers \cite{oneoJ} (see also \cite{Kruczenski:2004kw}). The work here could be useful in finding an 
expression for the one-loop corrections independent of the particular classical solution considered. 
This calculation of higher order terms in the effective action is interesting but will not be attempted here. 

 Another comment is about the cut-off. In the spin chain side there is a natural cut-off given by
the lattice spacing. In the string side although there is no cut-off in the original action, the 
reduced action also has a cut-off that determines its regime of validity. This follows already from the
BMN analysis which considers an ultrarelativistic string. If the momentum $J$ of the string is large and 
fixed, there is a maximum possible wave number for the string excitations, otherwise the mass of the string 
will be large and the ultrarelativistic approximation no longer valid. 

 The organization of this paper is as follows: in section \S\ref{reduced actions} we review
some properties of the reduced action and give the motivation for the calculation. In section
\S\ref{divergences} we study the action and analyze its scaling properties. 
In section \S\ref{one-loop} we use Wilson's renormalization group approach at
one loop to compute the flow of the action to the infrared. We end in section \S\ref{partcase} by giving
an example. Finally we give our conclusions in section \S\ref{conclu}.

\section{Reduced actions}
\label{reduced actions}

 In \cite{Kruczenski:2003gt} (see also \cite{Kruczenski:2004kw}), it was argued that the action
(which we write here in Euclidean space):
\beq
 S = - i \mu \int \cos\theta \dot{\phi} + \frac{\lambda}{2} 
           \int \left[(\partial_\sigma \theta)^2 + \sin^2\theta (\partial_\sigma \phi)^2 \right]
\label{SU(2)action}
\eeq
could be used to compute the anomalous dimensions of long operators in the $SU(2)$ sector of 
\N{4} SYM. The same action appears as a limit of the string action moving in the bulk. This action describes
the motion of spins pointing parallel to the direction $\vec{n}=(\sin\theta\cos\phi,\sin\theta\sin\phi,\cos\theta)$
subjected to a ferromagnetic interaction that tends to put them parallel to each other.
 Furthermore, $\mu$ is the spin of each site, namely it can be considered as a spin density, and $\lambda$ is a 
coupling constant that determines the strength of the interaction. The coordinate $\sigma$ is integrated from 
$0$ to the length of the chain $L$. This means that $\mu L$ is the maximum total spin, namely the spin of the 
ferromagnetic ground state.

This action can be put in the generic form (generalized to $d$ spatial dimensions):
\beq
 S = \mu \ints \left(\dot{z}^\mu\partial_\mu K -\dot{\bar{z}}^{\bmu}\partial_{\bmu}K\right)
 + \lambda \ints \partial_{\mu\bnu}K\, \partial_j z^{\mu}\partial_j \bar{z}^{\bnu}
\label{Kaction}
\eeq
 where $K$ is a K\"ahler potential in an $n$-complex dimensional K\"ahler manifold. Since in the rest of the
paper we study this action in detail it is appropriate to summarize here our notation. The sigma model is a field
theory in $d$ spatial dimensions denoted as $i,j,\ldots=1\ldots d$ and one Euclidean time denoted as $t$. Time
derivatives are denoted with a dot. The target manifold has complex dimension $n$ and holomorphic coordinates
$z^\mu$ with $\mu,\nu\ldots=1\ldots n$ and anti holomorphic $z^{\bmu}$ with $\bmu,\bnu,\ldots=1\ldots n$. We also find convenient at times 
to consider the manifold as a $2n$ dimensional manifold with coordinates $x^{\alpha}$ with $\alpha,\beta,\ldots = 1\ldots 2n$. 
 We also introduce a local frame (vielbein) with $n$ holomorphic one-forms $e^a_\mu$ labeled by $a,b,\ldots = 1\ldots n$ and 
$n$ anti-holomorphic $e^{\bar{a}}_{\bmu}$ labeled by $\bar{a},\bar{b}\ldots=1\ldots n$. We also consider a real 
frame $e^a_{\alpha}$ where now $a,b\ldots = 1\ldots 2n$. 

 As an example, (\ref{SU(2)action}) corresponds to $d=1$, $n=1$,  $K=\ln(1+z\bar{z})$,  
since a two sphere is a one dimensional complex manifold. 
 Another example appeared in \cite{BK} where the same type of action was found for a string moving fast in $\ads{5}\times Y^{p,q}$ in such a way
that each piece of the string moves approximately along a BPS geodesic.
 In the string side, the appearance of a K\"ahler metric is related to
supersymmetry but in the field theory side it seems a feature emerging from the coherent state method of
obtaining the classical action. In that case the complex manifold is a submanifold of $CP(n)$ which is the
space of states of a quantum system with $n$ discrete states. These states are the ones that can appear in 
a site of the spin chain. For example in the $SU(2)$ case there are two states and we get $CP(2)=S^2$. In general
however the states can be written in terms of a smaller set of parameters which parameterize the vacuum states.

 In the case of $Y^{p,q}$,  the target manifold of the sigma model has two complex dimensions 
and $SU(2)$ isometry. We refer the reader to \cite{BK} for the particular form of the K\"ahler potential since that 
is not essential for what we do in the present paper.
 To clarify a bit however we just mention that the four dimensional manifold in question is the base of the $Y^{p,q}$
manifold. The manifold $Y^{p,q}$ is five dimensional and can be written as a $U(1)$ fibration over the 
four dimensional K\"ahler base. The base has orbifold singularities although the 5-d manifold is regular. 

 The properties of the action (\ref{Kaction}), for generic $K$, is the focus of the rest of this paper.
 We should note that the actual action has an infinite number of terms from which we are considering only
the lowest order ones in an expansion in derivatives. This is allowed as long as we are concerned with the 
lowest energy states. From that point of view we study the flow under renormalization of these terms and
ignore the others.

\section{Perturbative expansion}
\label{divergences}

 In this section we analyze the action (\ref{Kaction}) to understand the generic properties of its
perturbative expansion. 
For this purpose it is convenient to consider the target manifold a $2n$ real manifold
(see below eqn.(\ref{Kaction}) for notation) and write the action as 
\beq
S = -i \mu \ints A_{\alpha} \dot{x}^{\alpha} + \frac{\lambda}{2} \ints g_{\alpha\beta} \partial_j x^\alpha \partial_j x^\beta
\eeq
 where $\alpha=\mu,\bmu$ and $A_{\mu}=i\partial_\mu K$, $A_{\bmu}=-i\partial_{\bmu} K$, $g_{\mu\bnu}=\partial_{\mu\bnu}K$. 
Since the action is adimensional we obtain the following units
\beq
 [\mu] = \frac{1}{L^d}, \ \ \ \ [\lambda]=\frac{1}{TL^{d-2}}
\eeq
where $L$ denotes units of length and $T$ of time. We see that $\mu$ is a density which in the $SU(2)$ case 
is the spin density and generically
we can call it a charge density. As we see below the theory is not renormalizable so we need to introduce a 
UV cutoff $\Lambda$ that we take to have
units of momentum ($[\Lambda]=1/L$). The only adimensional quantity we can construct is $\mu \Lambda^{-d}$. 
Since $\Lambda^{-1}$ is the minimal wavelength,
$\mu \Lambda^{-d}$ is the total spin or charge contained in the minimal volume we consider. For example, in 
the $SU(2)$ case, we can think that such 
elementary volume can be replaced by a single spin of value $S=\mu \Lambda^{-d}$. If $S\gg 1$ each spin (and 
therefore the whole system) behaves  
classically which leads us to expect that the loop counting parameter $\hbar$ is given 
by $\hbar=\frac{1}{S}=\frac{1}{\mu \Lambda^{-d}}$. This means that, 
as long as the waves we consider only move the spins in big groups, the classical approximation is valid. 

 Lets us make this more precise.  
Expanding $x^{\alpha}$  around a constant background $x^{\alpha}=x^{\alpha}_0+\delta x^{\alpha}$ it is easy 
to see that we can write a perturbative expansion 
with a propagator
\beq
\langle \delta x(k,\omega) \delta x (k,\omega) \rangle \sim \frac{1}{\lambda k^2+2i \mu\omega}
\eeq
and vertices of two types. One type of vertex is proportional to $\mu$ and contains one time derivative and
 the other is proportional to $\lambda$ and
contains two spatial derivatives. Both can contain arbitrary number of fields. A generic $n_l$-loop amplitude 
with $n_E$ external legs, 
coming from a Feynman diagram with $n^t_V$ vertices of the first type and $n^\sigma_V$ vertices of the second 
type is given schematically by:
\beq
A \sim \lambda^{n^\sigma_V} \mu^{n^t_V} \left[\int d^dk\, d\omega\right]^{n_l} 
 \left[\frac{1}{\lambda k^2+2i\mu\omega}\right]^{n_L}   
        \left(k^2\right)^{n^\sigma_V} \omega^{n^t_V} 
\eeq 
 where $n_L$ is the number of internal propagators and we suppressed the labels of the momenta. There are $n_l$ 
independent 
loop momenta which are integrated. Also the momenta of the propagators are linear combinations of the loop and 
external momenta. 
We can now rescale all $k$ and 
 $\omega$ as $k\rightarrow k/\sqrt{\lambda}$ , $\omega\rightarrow \omega/\mu$ to get
\beq
A \sim \frac{1}{\left(\mu\lambda^{\frac{d}{2}}\right)^{n_l}} \left[\int d^dk\, d\omega\right]^{n_l} 
 \left[\frac{1}{k^2+2i\omega}\right]^{n_L} \left(k^2\right)^{n^\sigma_V} \omega^{n^t_V} 
\eeq
 To do the integrals we put a cut-off such that $k^4+\omega^2 \leq \lambda^2 \Lambda^4 $ 
(or $\lambda^2 k^4+ \mu^2 \omega^2 \leq \lambda^2 \Lambda^4$ before the rescaling). 
If we rescale the external momenta with the cut-off, then, since $\sqrt{\lambda} \Lambda$ is the only remaining 
scale in the integral, we can use dimensional analysis to find that
\beq
A = \frac{1}{\left(\mu\lambda^{\frac{d}{2}}\right)^{n_l}} \left(\sqrt{\lambda}\Lambda\right)^{(d+2)n_l-2(n_L-n_V)}
  \Phi\left(\frac{p}{\Lambda},\frac{\mu}{\lambda}\frac{w}{\Lambda^2}\right) 
= \frac{\lambda \Lambda^2}{\left(\mu \Lambda^{-d}\right)^{n_l}} 
  \Phi\left(\frac{p}{\Lambda},\frac{\mu}{\lambda}\frac{w}{\Lambda^2}\right)
\eeq
where we used that the number of propagators $n_L$ is related to the number of vertices and loops by
\beq
 n_l = n_L-n^t_V-n^\sigma_V+1  
\eeq
by simple counting. We also wrote explicitly the dependence on the external momenta and energies $(p,w)$ through a function 
$\Phi$ that should be determined from the actual calculation. We assume here that the external momenta provide an infrared
cut-off to possible IR divergences. In the next section we use an alternative Wilsonian approach where one only integrates
a thin momentum shell and therefore no IR divergences can appear.      
 So, as expected, we obtain that the loop counting parameter is $\frac{1}{\mu\Lambda^{-d}}$ which is the inverse 
of the total charge 
or spin contained in the minimal volume we consider. When that is large, loop corrections are suppressed and the system 
behaves classically. In the effective action this amplitude contributes terms of the form 
\beq
S_{\mbox{eff}} \sim \frac{\lambda}{\left(\mu \Lambda^{-d}\right)^{n_l}} 
\left(\frac{\mu}{\lambda}\right)^{n_t} \Lambda^{2-2n_t-n_\sigma} 
    \ints h^{j_1\ldots j_{n_\sigma}}_{\alpha_1\ldots \alpha_{n_t},\beta_1\ldots \beta_{n_\sigma}} 
 \dot{x}^{\alpha_1}\ldots \dot{x}^{\alpha_{n_t}} 
 \partial_{j_1}x^{\beta_1}\ldots \partial_{j_{n_\sigma}}x^{\beta_{n_\sigma}}
\eeq
 where the coefficients $h^{j_1\ldots j_{n^\sigma}}_{\alpha_1\ldots \alpha_{n^t},\beta_1\ldots \beta_{n^\sigma}} $ can be
determined by expanding $\Phi$.  If the momenta are of order $k\sim \tilde{\Lambda}$ and the energies of order 
$\omega\sim \frac{\lambda}{\mu} \tilde{\Lambda}^2$ such term is of order 
\beq
 S_{\mbox{eff}}  \sim \frac{\lambda \Lambda^2}{(\mu\Lambda^{-d})^{n_l}} 
  \left(\frac{\tilde{\Lambda}}{\Lambda}\right)^{2n_t+n_\sigma}
\eeq
which expresses both, the fact that higher loop corrections are suppressed when $(\mu\Lambda^{-d})\gg1$ and
that higher derivative terms are suppressed at low energy ($\tilde{\Lambda}\ll \Lambda$). The first fact justifies 
the loop (or semi-classical) expansion and the second the restriction to lowest order terms.

\section{One-loop renormalization group}
\label{one-loop}

 In this section we study the one-loop renormalization group of the action.
 The renormalization of sigma models has been studied long ago by Honerkamp and collaborators~\cite{Honerkamp:1971sh}. 
The idea that the metric is modified by the loop corrections is well known from the work of Friedan \cite{Friedan:1980jm} 
and is a cornerstone of string theory (see \eg\ \cite{Callan:1985ia,Fradkin:1984pq}).  
 At the one-loop order it is convenient to use Wilson's procedure of integrating a thin 
momentum shell which was first applied to sigma models by Polyakov \cite{Polyakov:1975rr}.   

 As in the previous section, we write the action (\ref{Kaction}) in Euclidean space as
\beq
S = -i \mu \ints A_{\alpha} \dot{x}^{\alpha} 
  + \frac{\lambda}{2} \ints g_{\alpha\beta} \partial_j x^\alpha \partial_j x^\beta
\eeq
where $\alpha=\mu,\bar{\mu}$ runs over holomorphic and antiholomorphic indices and: 
\beqa
 A_\mu  &=& i\partial_\mu K  \\
 A_{\bmu}  &=& - i\partial_{\bar{\mu}} K  \\
 g_{\mu\bnu} &=& \partial_{\mu\bar{\nu}} K
\eeqa
 The action has a cut-off $\Lambda$ determining the maximum spatial momentum $k$ of the fields.
 To study the renormalization group we do three steps: 
\begin{itemize}
\item First we divide the fields into slow ($k<\Lambda e^{-b}$) and fast  
($\Lambda e^{-b}<k<\Lambda$) and integrate out the fast modes getting a theory with cut-off $\Lambda e^{-b}$.
\item Then we rescale the coordinates so that the cut-off goes back to being $\Lambda$. 
\item Finally we rescale $K$ to normalize the target metric.
\end{itemize}

 For the first step we should do the separation in fast and slow modes in a generally covariant way. This is not only
to get more elegant expressions but also because, in general, to parameterize the target manifold, we need to introduce 
different coordinate patches. Coordinates in different patches are related by a coordinate transformation and therefore,
if our procedure is not invariant under those, the resulting action will not be well defined. In the case of a complex
manifold we are interested in, we only need to ask for invariance under holomorphic transformations.  
 
 The best way to do the expansion is to use normal coordinates~\cite{Honerkamp:1971sh,Friedan:1980jm} and consider fluctuations as
\beq
x^\alpha= x_0^\alpha + \xi^\alpha - \half \Gamma^{\alpha}_{\beta\gamma} \xi^{\beta}\xi^{\gamma} 
         + \frac{1}{6} \left(2\Gamma^{\alpha}_{\beta\gamma} \Gamma^{\beta}_{\phi\delta} - \partial_{\gamma}\Gamma^{\alpha}_{\phi\delta}\right)
         \xi^{\gamma} \xi^{\delta} \xi^{\phi} + \ldots
\eeq
where $\Gamma^{\alpha}_{\beta\gamma} = \half g^{\alpha\alpha'}\left(g_{\beta\alpha',\gamma}+g_{\gamma\alpha',\beta}-g_{\beta\gamma,\alpha'}\right)$
is the connection. Expanding a generic tensor we get (here we follow \cite{Alvarez-Gaume:1981hn})
\beq
\frac{\partial x^{\alpha_1}}{\partial\xi^{\beta_1}} \ldots \frac{\partial x^{\alpha_l}}{\partial\xi^{\beta_l}} T_{\alpha_1\ldots\alpha_l} = 
T_{\beta_1\ldots \beta_l} + T_{\beta_1\ldots\beta_l;\alpha} \xi^{\alpha} + 
\half \left\{ T_{\beta_1\ldots \beta_l;\alpha\beta}+\frac{1}{3} \sum_i R^{\alpha_i}{}_{\alpha\beta\beta_i} T_{\beta_1\ldots\alpha_i\ldots \beta_l} \right\} 
\xi^\alpha \xi^\beta + \ldots
\eeq
where $R^{\alpha}{}_{\beta\gamma\delta}$ is the Riemann tensor and the semicolon ``;'' indicates covariant derivative.
In particular
\beq
\frac{\partial x^\beta}{\partial\xi^\alpha} A_\beta = A_{\alpha} + A_{\alpha;\beta} \xi^{\beta} + \half\left\{A_{\alpha;\beta\gamma} 
          + \frac{1}{3} R^{\delta}{}_{\beta\gamma\alpha} A_\delta \right\} \xi^\beta\xi^\gamma +\ldots
\eeq
Also we get from \cite{Alvarez-Gaume:1981hn}
\beq
\frac{\partial \xi^{\beta}}{\partial x^\alpha} \partial_j x^\alpha = \partial_j x_0^\beta + D_j \xi^{\beta} 
  + \frac{1}{3} R^{\beta}{}_{\gamma\delta\alpha}\, \partial_j x_0^{\alpha}\, \xi^{\gamma} \xi^{\delta} +\ldots
\eeq
where we defined the covariant derivative
\beq
D_j \xi^{\beta} = \partial_j \xi^{\beta} + \Gamma^{\beta}_{\alpha\gamma}\, \partial_jx_0^\alpha\, \xi^{\gamma}
\eeq
 Finally, we introduce a set of vielbeins $e^{\alpha}_a(x_0)$ such that $\delta^{ab} e^{\alpha}_a e^{\beta}_b = g^{\alpha\beta}$ and
fluctuations $\xi^a$ through 
\beq
\xi^\beta = e^{\beta}_a\, \xi^a
\eeq
 which leads to the covariant derivative
\beq
D_j \xi^a = \partial_j \xi_a + \omega_\alpha{}^a{}_b \,\partial_j x^{\alpha} \, \xi^b
\eeq
Here $\omega_\alpha{}^a{}_b$ is the usual spin connection defined as
\beq
\omega_\alpha{}^a{}_b = e^a_\beta\, e^\beta_{b;\alpha}
\eeq
 Since the vielbein is arbitrary, the classical action has an $SO(2n)$ gauge invariance. Again, since usually a vielbein is not 
globally defined, 
the perturbation procedure should also be $SO(2n)$-gauge invariant so that one can transition between different 
patches with vielbeins
differing by arbitrary rotations. In our case however we need only to respect a $U(n)$ gauge invariance since 
the vielbeins are 
divided into holomorphic and antiholomorphic and we consider only rotations that do not mix them. 

Putting all together we expand the action to second order as
\beq
S = S^{(0)} +S^{(2)}
\eeq
with
\beqa
S^{(2)} &=& \frac{i\mu}{2}\int F_{ab} \xi^b D_t\xi^a  + \frac{i\mu}{2} \int F_{\alpha a;b} \xi^a\xi^b \dot{x}_0^\alpha 
                                                                                                   \nonumber \\
&& + \frac{\lambda}{2} \int D_j\xi^aD_j\xi^a + \frac{\lambda}{2} \int R_{\alpha a b \beta}\, 
                        \partial_j x_0^\alpha \partial_j x^\beta_0\, \xi^a\xi^b
\eeqa
 where $F_{\alpha\beta}= \partial_\alpha A_\beta - \partial_\beta A_\alpha$ is the field strength associated 
with $A_\alpha$ and $F_{ab} = e^\alpha_a e^\beta_b F_{\alpha\beta}$. 
 We note that the first order term in the action vanishes since there is no overlap between fast and slow modes.

To proceed we are going to consider the particular case we are interested in where
\beq
 A_\mu = i \partial_\mu K,\  A_{\bmu} = -i \partial_{\bmu} K\ \ \  \Rightarrow\ \ \  F_{\mu\bnu} = -2i \partial_{\mu\bnu} K = -2i g_{\mu\bnu} 
\label{AFf}
\eeq
 This means that
\beq
F_{a\bar{b}} =  -2i\delta_{a\bar{b}} 
\eeq
where we split $a$ into holomorphic and antiholomorphic indices $a, \bar{a}$ (see below eq.(\ref{Kaction}).
Furthermore, from (\ref{AFf}) we obtain:
\beq
F_{\alpha a;b}=0
\eeq
as we knew since $F_{\alpha\beta}$ is actually the K\"ahler form.
 Using this, the quadratic action reduces to
\beqa
S^{(2)} &=& 2\mu \int \xi^{\bar{a}}\partial_t\xi^{a} 
               + 2\mu \int \omega_{\alpha\bar{a}b}\xi^{\bar{a}}\xi^{b}\dot{x}_0^\alpha 
        +\lambda\int D_j \xi^{\bar{a}} D_j\xi^a \\
        && + \frac{\lambda}{2}\int 
           \left(R_{\mu a\bar{b}\bnu} + R_{\bnu a\bar{b}\mu}+R_{\mu \bar{b}a\bnu}+R_{\bnu \bar{b}a\mu}\right) 
           \partial_jx_0^\mu \partial_jx_0^{\bnu} \xi^a\xi^{\bar{b}} 
\eeqa
 To proceed we should introduce a cut-off in a gauge invariant way. For example we can use a proper time 
regularization~\cite{Schwinger:1951nm} to define the inverse and determinant of the quadratic operator. 
From a more modern perspective we can probably use dimensional regularization in the spatial directions 
although it is not clear then how to treat the time direction. After doing the gauge invariant regularization, 
the spin connection $\omega_\alpha{}^a{}_b$, which is the $U(n)$ gauge field, can appear only in gauge invariant 
combinations. If we call 
its field strength (which is esentially the Riemann tensor) $\cF$ then, due to rotational invariance, at first sight 
we can only have combinations such as $\cF_{0j} \cF_{0j}$ or $\cF_{ij} \cF_{ij}$ 
which are higher order in time or spatial derivatives. However, one can see that a possible gauge invariant term in the 
action is $\int \omega_\alpha{}^a{}_a\, \dot{x}^{\alpha}$ which is gauge invariant since $\omega_\alpha{}^a{}_a$
is a $U(1)$ gauge field, corresponding to the $U(1)$ factor in the holonomy group $U(n)=SU(n)\times U(1)$ (where n is 
the complex dimension of the manifold). 
 In a general manifold we can attempt to consider a similar term $\int \Gamma^{\alpha}_{\beta\alpha}\, \dot{x}^\alpha$ 
but that fails  since it is a total derivative $\int \partial_t \ln g =0$. The term we discussed is invariant because we restrict 
ourselves to holomorphic transformations.  
 
 Therefore, at one loop, using a gauge invariant regularization,  is equivalent to replacing $D_j \rightarrow \partial_j$ 
in the third term of $S^{(2)}$. From the $D_t$ part however we get the second term whose expectation value is 
precisely the gauge invariant term that we have just discussed. 

 Now we do a Fourier transform (we assume the background fields are approximately constant from the point of 
view of the fast variables $\xi^a$): 
\beq
\xi^a(x,t) =\int\frac{d^dk}{(2\pi)^d}\frac{d\omega}{2\pi}\, \xi^a(k,\omega)\, e^{ikx+i\omega t}
\eeq
and get the $\xi^a$ propagator
\beq
 \langle \xi^{\bar{a}}(k_1,\omega_1) \xi^b(k_2,\omega_2) \rangle = \frac{(2\pi)^{d+1}}{\lambda k_1^2+2i\mu\omega_1} \delta^{\bar{a}b}
                                           \delta(k_1-k_2)\delta(\omega_1-\omega_2)
\eeq
 With this we compute $\langle S^{(2)}\rangle $. At this point we get delta functions evaluated at zero in momentum space
that can be interpreted as spatial and time integrations of the background fields, namely $\delta_k(0)\delta_w(0)=\int \frac{d^dx\,dt}{(2\pi)^{d+1}}$. 
The result is
\beq
\langle S^{(2)}\rangle  = 2\mu\cC \ints \omega_{\alpha}{}^a{}_a\, \dot{x}_0^{\alpha} 
                          - \lambda \cC \ints R_{\mu\bnu}\, \partial_j x_0^\mu \partial_jx_0^{\bnu} 
\eeq
where $\cC$ is the integral over a thin momentum shell ($b\rightarrow 0^+$):
\beq
\cC = \int_{\Lambda e^{-b}}^\Lambda \frac{d^d k d\omega}{(2\pi)^{(d+1)}} \frac{1}{\lambda k^2+2i\mu\omega}
  \simeq_{b\rightarrow 0} \frac{1}{(4\pi)^{\frac{d+1}{2}}} \frac{1}{2\mu\Lambda^{-d}} 
    \frac{\Gamma(\frac{2+d}{4})}{\Gamma(\frac{d}{2})\Gamma(1+\frac{d}{4})} \ b = \bar{\cC} b
\eeq
 is the momentum integral and we defined $\bar{\cC}=\lim_{b\rightarrow 0} (\cC/b)$. The cut-off was introduced through
\beq
 (\lambda^2k^4 + \mu^2\omega^2)\le \lambda^2\Lambda^4
\eeq
 so that $\Lambda$ is interpreted as a momentum cut-off ($k\le \Lambda$). The energy cut-off is 
$\omega\le \frac{\mu}{\lambda}\Lambda^2$. As expected we see the correction to be of order $1/(\mu\Lambda^{-d})$
as we discussed in the previous section.

 Therefore, ignoring all terms higher order in derivatives (since we are considering a low energy expansion of the 
effective action), the one-loop terms effectively shift the metric and gauge field to
\beqa
 \tilde{g}_{\mu\bnu} &=& g_{\mu\bnu} - \cC R_{\mu\bnu} \\
 \tilde{A}_{\mu} &=& A_{\mu} + 2i\,\cC\, \omega_{\mu}{}^a{}_a \\
 \tilde{A}_{\bnu} &=& A_{\bnu} - 2i\,\cC\, \omega_{\bmu}{}^{\bar{a}}{}_{\bar{a}} 
\eeqa
 Now we use that
\beq
 R_{\mu\bnu} = -\partial_{\mu\bnu} \ln \det g_{\rho\bar{\sigma}} 
\eeq
 and
\beq
 \omega_\alpha{}^a{}_a = e^a_\nu \, e^\nu_{a;\alpha} = e^a_\nu\,\partial_\alpha e^\nu_a + \Gamma^\beta_{\alpha\beta}  =
        - i \partial_\alpha \phi + \half \partial_\alpha \ln \det  g_{\rho\bar{\sigma}}
\eeq
 where $\phi$ in an undetermined phase that depends on the choice of vielbein through $\det e^a_\nu = e^{i\phi}\sqrt{\det g_{\rho\bar{\sigma}}}$. 
Fortunately, when replacing in the action, $\phi$ appears in a total derivative: $\int \partial_t \phi =0$ 
(actually this is the statement that such term is $U(1)$ gauge invariant). 
 Using now the fact that $A_\mu=i\partial_\mu K$ and $g_{\mu\bnu}=\partial_{\mu\bnu} K $ we see that 
the action has exactly the same form (\ref{Kaction}) but with a K\"ahler potential
\beq
 \tilde{K} = K + \cC \ln\det g_{\rho\bar{\sigma}}
\eeq

Now we go to the second step and rescale the coordinates $x_j\rightarrow e^b x_j$ and $t\rightarrow e^{2b} t$ in order to restore 
the cut-off to its original value $\Lambda$ (namely $k\rightarrow e^{-b}k$ so the new $k$ extends to $\Lambda$ instead of $e^{-b}\Lambda$). 
This is the standard cut and stretch procedure that zooms into the low energy region. From the action we see
that this just rescales $K$ by $e^{bd}$. So we get
\beq
 \tilde{K} = K + b \left( \bar{\cC} \ln\det g_{\rho\bar{\sigma}} + K d\right)
\eeq
 where $\bar{\cC}=\cC/b$, \ie\ we extract the $b$ factor. 
   We can then write a $\beta$-function as
\beq
 \beta_K =  \partial _b K = \bar{\cC} \ln\det g_{\rho\bar{\sigma}} +  K d
\eeq
 In terms of the metric we have
\beq
 \partial_b g_{\alpha\beta} = - \bar{\cC} R_{\alpha\beta} + d\, g_{\alpha\beta}
\label{gflow1}
\eeq
 For a relativistic sigma model this type of equation is well-known. The only point we needed here is that the
term linear in time derivatives also renormalizes in the same way and the action has an invariant form\footnote{We should
note also that the relativistic calculation is special in two dimensions since then the classical action is
scale invariant and the second step, namely rescaling the coordinates, does not change the metric. Therefore, 
the second term on the right hand side of (\ref{gflow1}) is absent in that case.}.
 This equation has an Einstein metric as a fixed point. If we are away from the fixed point, the metric flows. 
The first thing to note is that the volume changes under such flow since:
\beq
 \partial_b \sqrt{g} = \half \sqrt{g} g^{\alpha\beta} \partial_b g_{\alpha\beta} 
   = \half \sqrt{g} \left(-\bar{\cC} R + d(2n)\right) 
\eeq
 where $R$ is the Ricci scalar. For the volume $V=\int\sqrt{g}$, we get
\beq
 \frac{1}{V} \partial_b V = \half \left( -\bar{\cC} r + d(2n)\right), \ \ \ \mbox{with}\ \ \  r=\frac{1}{V}\int \sqrt{g}R
\eeq
Then, as a final step, we can rescale $K$ in such a way that the volume is fixed.
Rescaling $K\rightarrow e^{\chi b} K$ with $\chi=\frac{r}{2n} \bar{\cC} - d $ the flow equation becomes
\beq
 \partial_b g_{\alpha\beta} = \bar{\cC}\left( - R_{\alpha\beta} + \frac{r}{2n}\, g_{\alpha\beta} \right)
\label{gflow2}
\eeq  
 which is the so-called normalized Ricci flow. In the case of a compact K\"ahler manifold, under this flow, 
the metric is known to flow to a K\"ahler-Einstein metric~\cite{Cao}. However, we are interested also in the 
non-compact case of which we consider an example in the next section.

Finally, to compensate for the rescaling of $K$ we have to rescale $\mu$ and $\lambda$ getting
\beqa
 \beta_\mu = -\chi \mu \\
 \beta_\lambda = -\chi \lambda
\eeqa
 We see that the ratio $\mu/\lambda$ is fixed. We only change an overall scale in the action. Since $\bar{\cC}$ is
small where the one-loop approximation is valid we have $\chi<0$. That means that the overall factor in the action grows
and the system becomes classical. This is just the naive idea that as we lower the cut-off there are more
spins in the minimal volume and therefore the system is more classical. The one-loop result is a small correction.

 To summarize, the final result is that as we lower the cut-off the system becomes more classical and the metric flows to 
an Einstein metric. 

 Note that in the $SU(2)$ case (\ref{SU(2)action}) the metric of the sphere is already Einstein so nothing happens except that
the system becomes classical in the infrared as follows from the simple reasoning already explained. However, 
in general the problem remains non-trivial. For example if we would want to compute the finite temperature 
partition function, then, at low temperature we need the partition function of a classical ferromagnet
which is a non-trivial problem already studied for example by Polyakov using Wilson's 
renormalization group in \cite{Polyakov:1975rr}. However, in the case of interest for the reduced string action we have
$d=1$ and the classical partition function can actually be computed simply by considering the problem as the evaluation
of a quantum mechanical propagator where the eigenstates are spherical harmonics. The result is
\beq
Z_{\mbox{cl.}}(\beta) = \int \cD\vec{n} e^{-\frac{\beta\lambda}{2}\int_0^L (\partial_\sigma\vec{n})^2\, d\sigma}
 = \sum_{\ell=0}^{\sqrt{\beta\lambda\Lambda}} (2\ell+1)e^{-\frac{L}{\beta\lambda}\ell(\ell+1)}  
\eeq
 From the field theory point of view, this partition function determines the distribution of the lowest anomalous dimensions 
of very long operators in the $SU(2)$ sector. 

\section{A particular case}
\label{partcase}

 When studying strings propagating in $\ads{5}\times Y^{p,q}$ a sigma model of the type we consider in this paper appears.
The complex manifold has two complex dimensions $z_{1,2}$ with an $SU(2)$ isometry. The K\"ahler potential 
depends on $\rho=\bar{z}_1 z_1 + \bar{z}_2 z_2$.
 Parameterizing the complex coordinates as
\beqa
 z_1 &=& \sin\frac{\theta}{2} \, e^{-\frac{i}{2}(\beta-\phi)} \sqrt{\rho} \\
 z_2 &=& \cos\frac{\theta}{2} \, e^{-\frac{i}{2}(\beta+\phi)} \sqrt{\rho} 
\eeqa
 the action becomes
\beqa
 S &=& i \mu \int d\sigma dt K_x\left(\dot{\beta} +\cos\theta \dot{\phi}\right) \\
   && + \frac{1}{4} \lambda \int d\sigma 
      dt \left\{ K_x \left((\partial_\sigma \theta)^2 + \sin^2\theta (\partial_\sigma \phi)^2 \right)
      + K_{xx} dx^2 + K_{xx} \left(\partial_\sigma \beta+\cos\theta \partial_\sigma \phi\right)^2\right\} \nonumber
\eeqa 
 where it turned out to be convenient to use $x=\ln \rho$ as a coordinate. From the positivity of the metric we 
see that we need $K_x>0$ and $K_{xx}>0$.  We can now compute
\beq
 \det g_{\rho\bar{\sigma}} = e^{-2x} K_x K_{xx}
\eeq
 and therefore the RG equation is
\beq
\partial_b K(x) = \bar{\cC} \left(-2x + \ln K_x + \ln K_{xx} + \frac{d}{\bar{\cC}} K \right) 
\eeq 
 where we used eq.(\ref{gflow1}) since now the manifold is non-compact and it is not clear that normalizing
the volume is appropriate.
 In the case studied in \cite{BK}, the manifold has orbifold singularities at $\rho=0,\infty$. Therefore we 
are in the non-compact case and no rigorous mathematical result applies (at least that we know of). The type 
of singularity is such that $K$ behaves as
\beq
 K \simeq_{x\rightarrow \pm \infty} \eta_{\pm} x + A_{\pm} e^{\alpha_{\pm}x} + \cO\left(e^{2\alpha_{\pm} x}\right)
\eeq
 Using this in the RG equation we can see that $\alpha_{\pm}$ are not renormalized. The slopes $\eta_\pm$
however, change according to:
\beq
\partial_b \eta_{\pm} = \bar{\cC} \left(\alpha_{\pm}-2+ \frac{d}{\bar{\cC}}\eta_{\pm}\right)
\eeq
 The fixed point is at $\eta_{\pm} = -\frac{\bar{\cC}}{d} (\alpha_{\pm}-2)$. 
The problem is, however, that if the slopes $\eta_{\pm}$ are not at the fixed point then they flow away from 
it. This means that we have to start with the correct value of $\alpha_{\pm}$ and $\eta_{\pm}$. By fixing the 
metric at the end points, we expect that the system behaves as in the compact case and flows to the 
K\"ahler-Einstein metric although we did not verify that explicitely. 
 
 If we study the simplified model proposed in \cite{BK} it turns out that the constraints on the slopes are 
not satisfied\footnote{It should be noted however that in those cases the volume is finite even if the 
manifold has singularities so we also expect the system to be similar to the compact case after we impose 
conditions near the singularities.}. This means that the model, or at least the way in which the continuum 
limit was taken, was perhaps too naive.  

 The general idea however seems correct in the sense that an infinite class of models flow to the IR fixed 
point implying that most of the details of the gauge theory disappear. This is specially true in the compact case. 
Therefore we conclude that the (classical) string action captures only the main features of the dilatation operator. 

\section{Conclusions}
\label{conclu}

 Motivated by some ideas described in \cite{BK}, in this paper we studied the one-loop renormalization of certain actions
given in terms of a K\"ahler potential $K$. We showed that, as expected, when we lower the cut-off, the system
behaves more classically and $K$ flows to a K\"ahler-Einstein metric. 
This follows from a known mathematical result for K\"ahler-Ricci flows on compact manifolds. 
In the case of the reduced action, non-compact cases also appear and we suggested that
by putting appropriate boundary conditions, the same result should follow. However a more detailed 
analysis of this point is desirable. 
  
 One point to emphasize is that we consider the action from a low energy effective action point of view. The action
has infinite number of terms and we consider only the lowest ones in an expansion in derivatives. In this sense
we study the renormalization group of only those terms and ignore the others which are irrelevant (at least in 
the case $d=1$ that we are interested in).

 From a generic perspective, this action computes the anomalous dimensions of long operators in the large-N 
limit of a field theory. We see that, when the operators are very long, and we concentrate in the lowest
anomalous dimensions, the effective action is determined by a fixed point and therefore is largely independent
of the details of the theory. The hope is then that the large N-limit of a gauge theory could be independent
of a detailed evaluation of the planar diagrams and should be given by statistical considerations (at least
when dealing with large number of fields or particles).

\section{Acknowledgments}

 I am grateful to S. Benvenuti and A. Tseytlin for collaboration on related projects. 
 I am also indebted to G. Ferretti, J. Maldacena, D. Martelli and A. Polyakov for various 
comments. 

This material is based upon work
supported by the National Science Foundation Grant No.
PHY-0243680.
Any opinions, findings, and conclusions or recommendations expressed in
this material are those of the authors and do not necessarily reflect
the views of the National Science Foundation.



\end{document}